 \documentclass[aps,pra,superscriptaddress,amsmath,amssymb,preprintnumbers,twocolumn,floatfix,showpacs,showkeys,10pt]{revtex4-1}
 \usepackage{amssymb} \usepackage{epsfig}
 \begin{document}
  \title{Preservation of quantum Fisher information and geometric phase of a single qubit system in a dissipative reservoir through the addition of qubits}

\author{You-neng Guo}
\affiliation{Department of Electronic and Communication Engineering, Changsha University, Changsha, Hunan
410022, People's Republic of China}
\affiliation{Key Laboratory of Low-Dimensional Quantum Structures and
Quantum Control of Ministry of Education, and Department of Physics,
Hunan Normal University, Changsha 410081, People's Republic of
China}
\author{Tian Qing-long}
\email{chinatql@126.com}
\affiliation{Department of Mathematics and Computing Science, Changsha University, Changsha, Hunan
410022, People's Republic of China}
\author{Mo Yun-fei}
\affiliation{Department of Electronic and Communication Engineering, Changsha University, Changsha, Hunan
410022, People's Republic of China}
\author{Ke Zeng}
\email{zk92@126.com}
\affiliation{Department of Electronic and Communication Engineering, Changsha University, Changsha, Hunan
410022, People's Republic of China}

\begin{abstract}
In this paper, we have investigated the preservation of quantum Fisher information (QFI) of a single-qubit system coupled to a common zero temperature reservoir through the addition of noninteracting qubits. The results show that, the QFI is completely protected in both Markovian and non-Markovian regimes by increasing the number of additional qubits. Besides, the phenomena of QFI display monotonic decay or non-monotonic with revival oscillations depending on the number of additional qubits $N-1$ in a common dissipative reservoir. If $N < N_{c}$( a critical number depending on the reservoirs¡¯ parameters), the behavior of QFI with monotonic decay occurs. However, if $N \geq N_{c}$, QFI exhibits non-monotonic behavior with revival oscillations. Moreover, we extend this model to investigate the effect of additional qubits $N-1$ and the initial conditions of the system on the geometric phase (GP). It is found that, the robustness of GP against the dissipative reservoir has been demonstrated by increasing gradually the number of additional qubits. Besides, the GP is sensitive to the initial parameter $\theta$, and possesses symmetric in a range regime $[0,2\pi]$.
 \end{abstract}

 \maketitle
\section{Introduction}
Quantum Fisher information (QFI) which extends the classical Fisher Information plays a significant role in the fields of quantum metrology~\cite{1}.  QFI characterizes the sensitivity of the state with respect to changes in a parameter~\cite{2}.  According to the quantum estimation theory,
the ultimate achievable parameter estimation precision is characterized by the QFI through the Cram$\acute{e}$r-Rao inequality~\cite{2a,2b}. For
an estimation parameter with a larger QFI value, the accuracy is more clearly achieved. Moreover, QFI may serve as a new resource
in quantum information tasks to witness entanglement detection~\cite{3}, non-Markovianity characterization~\cite{4}, and uncertainty relations~\cite{5,6} and so on.

However, any realistic quantum system of interest is unavoidably disturbed by surrounding environments which not only result in the loss of quantum coherence of interest systems but also degrease of parameter estimation precision. Hence, how to preserve and enhance the QFI becomes a key problem to be solved. In recent years, great attention has been paid to protecting and improving the QFI. Different protocols and strategies
have been proposed and realized in the quantum metrology.  For instance,  Tan et al. investigated the improvement precision of parameter estimation in atom interferometer suffering from dephasing noise by using dynamical decoupling pulses~\cite{Tan1,Tan2}. Zhang et al. proposed an alternative scheme to enhance the precision of quantum estimation by introducing the quantum correlations of the initial states~\cite{Zhang}. By employing external field driven, Abdel-Khalek studied the QFI for a two-level atom system in a dissipative environment~\cite{Abdel-Khalek}. Inspired by the classical driving, Li et al. investigated parameter estimation precision of a
non-Markovian dissipative two-state system~\cite{Li1}. Taking non-Markovian effect into consideration, a qubit system in non-Markovian environment becomes a good candidate for implementation of quantum optics schemes and information with high precision~\cite{Berrada1, Berrada2}.  Zheng et al. analyzed the parameter precision of optimal quantum estimation of a dissipative qubit by applying quantum feedback technology~\cite{Zheng}. Other schemes such as, for protecting the QFI to enhance parameter estimation precision are made use of weak measurement and quantum measurement reversal~\cite{He, Xiao}. In this paper, we are going on to preserve and improve QFI in a dissipative environment using addition noninteracting qubits. The consideration of the additional qubits to suppress decoherence was first observed in~\cite{An}, where the bipartite entanglement is protecting by the addition of a third qubit. Later, many more additional qubits are introduced to fight against decoherence~\cite{An1,Flores,Behzadi,Behzadi1}. For example, Behzadi et al. investigated quantum coherence preservation of a two-level system in a dissipative reservoir via the addition of qubits into the reservoir~\cite{Behzadi}.

On the other hand, the phase of quantum states which is a fundamental property of quantum coherence, can be measured by interfering the wave function
with another coherent wave function so that one can discern whether or not the system has undergone an evolution. The phase that depends only on the geometry of the path traversed by the system during its adiabatic evolution, is first named the geometric phase (GP) by Berry~\cite{Berry}. GP has observable consequences in a wide range of systems~\cite{Tomita,Leek,Chen}. It can not only be used to encode information on systems~\cite{Jones}, but also treated as a quantifier to detect quantum criticality in the systems of spin chain.~\cite{Yi,Zhang2}. In resent years, the topic of the GP has recently been studied intensively and is drawing increasing
attention. On one hand, it has been generalized from the cases of adiabatic, cyclic and unitary evolutions to the cases of non-adiabatic, non-cyclic and nonunitary evolutions~\cite{Aharonov,Samuel,Tong}. On the other hand, it is aslo generalized from the cases of pure-state or  nondegenerate density operators undergoing unitary evolutions~\cite{Sjoqvist} to the cases of mixed-state or degenerate density operators in nonunitary evolutions~\cite{Uhlmann,Singh, Ericsson,Peixoto}. Besides, another direction of the generalization is to study the composite system GP, especially one considers the relation of GP between the composite system and its subsystems. In general, the GP of the composite system in nonlocal unitary evolution is always not equal to the sum of the GP of its subsystems~\cite{Tong1,Williamson}, except for the case where the composite system with initial separable state undergoes local unitary evolution~\cite{Niu}. Among these extensions of the GP, one based on mixed-state density operators undergoing nonunitary evolutions has been extensively studied in various contexts.

The purpose of this paper is going on to examine the protection of QFI and GP from decoherence environment. To do so, we will consider an N-qubit system embedded in a common reservoir which consists of harmonic oscillators. The model under investigation is simple and has been drawing increasing attention in various contexts due to its with an exact analytical solution under some assumption~\cite{Flores,Behzadi,Behzadi1}. In this paper, our main motivation is mainly concentrated on how the additional qubits affect QFI of phase estimation and GP. Specifically we compare the behaviors of QFI in both weak and strong coupling regimes through the addition of noninteracting qubits. The results show that, the dynamics of QFI displays monotonic decay or non-monotonic with revival oscillations depending on the number of additional qubits $N-1$ in a common dissipative reservoir. If $N < N_{c}$, the behavior of QFI with monotonic decay occurs, the larger the number of additional qubits, the more slowly the QFI decays and otherwise, QFI exhibits non-monotonic behavior with revival oscillations, with the number of additional qubits  increasing, the QFI is protected well. Furthermore, we calculate the GP for a nonunitary evolving qubit according to the formula of Ref.~\cite{Tong}, and investigate the effect of the addition qubits and initial conditions of the system on the geometric phase (GP). It is found that, the GP is immune to the decoherence of environment in both Markovian and non-Markovian regimes when the number of additional qubits is large enough, but it is sensitive to the initial parameter $\theta$, and possesses symmetric in a range regime $[0,2\pi]$.

This paper is organized as follows: In Sec.\textrm{2}, we introduce the model based on an $N$-qubit system immersed in a common zero-temperature thermal reservoir and its analytical solution. In Sec.\textrm{3}, the protection of QFI for a single qubit system coupled to a common reservoir through the addition of qubits is investigated.  In Sec.\textrm{4}, the influence of additional qubits as well as initial condition on GP is studied. Finally, we give the conclusion  in Sec.\textrm{5}.

\section{The model and its solution}  %%% ½Ú±êÌâ 2
We consider a single-qubit system along with $N-1$ similar non-interacting qubits located in a common dissipative reservoir. For simplicity, we suppose that $N$ non-interacting qubits are independently coupled to a common zero temperature thermal reservoir which is consistent of harmonic oscillators. The Hamiltonian of the whole system is given by~\cite{Behzadi,Behzadi1},
\begin{equation}
H=\omega_{0}\sum_{i=1}^{N}\sigma_{i}^{+}\sigma_{i}^{-}+\sum_{k}\omega_{k} b_{k}^{\dagger}b_{k}+\sum_{i=1}^{N}\sum_{k}(B^{\dagger}\sigma_{i}^{-}+B\sigma_{i}^{+}),
\end{equation}
where $B=\sum_{k} g_{k}b_{k}$,   $g_{k}$ are the coupling strength between the $i^{th}$ and field mode $k$ in the reservoir,  $b_{k}^{\dagger}$ and
$b_{k}$ are the mode $k$ creation and annihilation operators, respectively. $\omega_{k}$ donates  the $k_{}^{th}$ mode with frequency in the reservoir,
$\sigma_{i}^{+}=|1\rangle_{i}\langle0|$ and $\sigma_{i}^{-}=|0\rangle_{i}\langle1|$ are the raising and lowering operators of the $i^{th}$ qubit
system with the same transition frequency $\omega_{0}$, respectively.

For simplicity, we consider the situation of no more than one excitation in the whole system, and the reservoir is initially in the vacuum state$|0\rangle_{E}$. Supposing the initial state of the N-qubit system has the form $|\psi(0)\rangle=C_{0}(0)|0\rangle_{s}+\sum_{i=1}^{N}C_{i}(0)|i\rangle_{s}$, where $|0\rangle_{s}$ donates all of the qubits are in the ground state $|0\rangle$, and $|i\rangle_{s}$ donates the $i^{th}$ qubit state is in the excited state $|1\rangle$ and the rest ones are in the ground state $|0\rangle$. Then the total system state $|\Phi(0)\rangle=|\psi(0)\rangle\otimes|0\rangle_{E}$ evolves after time $t > 0$ into the state $|\Phi(t)\rangle=[C_{0}(t)|0\rangle_{s}+\sum_{i=1}^{N}C_{i}(t)|i\rangle_{s}]\otimes|0\rangle_{E}+\sum_{j=1}^{N}C_{j}(t)|0\rangle_{s}\otimes|1_{j}\rangle_{E}$. The equality $|C_{0}(t)|^2+\sum_{i=1}^{N}|C_{i}(t)|^2+\sum_{j=1}^{N}|C_{j}(t)|^2$ holds due to the normalization condition. Similarly, we denote $|0\rangle_{E}$ being the vacuum state of the reservoir and $|1_{j}\rangle_{E}$ the state of it with only one excitation in the $j^{th}$ field mode. By solving the Schrodinger equation in the interaction picture, the reduced density matrix of the $i^{th}$ qubit is obtained by partial tracing out both the environmental reservoir's degrees of freedom and the other qubits
\begin{equation}\label{E1}\rho_{i}(t)=\left(
\begin{array}{ c c c c l r }
\rho_{11}^{i}(0)|C_{i}(t)|^2 & \rho_{10}^{i}(0)C_{i}(t)\\
\rho_{01}^{i}(0)C_{i}^{\ast}(t) & \rho_{00}^{i}(0)+\rho_{11}^{i}(0)(1-|C_{i}(t)|^2)\\
\end{array}
\right),
\end{equation}
where $\rho_{11}^{i}=|C_{i}(0)|^2$, $\rho_{00}^{i}=|C_{0}(0)|^2$, and $\rho_{10}^{i}=\rho_{01}^{i\ast}=C_{i}(0)C_{0}^{\ast}(0)$.
The function $C_{i}(t)$  satisfies
\begin{equation}\label{E2}
\frac{d}{dt}C_{i}(t)=-\int_{0}^{\infty}\int_{-\infty}^{+\infty}J(\omega)e^{i(\omega_{0}-\omega_{k})(t-\tau)}\sum_{j=1}^{N}C_{j}(\tau)d\omega d\tau,
\end{equation}
where $J(\omega)$ is the spectral density function, characterizing the reservoir
spectrum. To solve the function $C_{i}(t)$, we assuming the reservoir as a Lorantzian spectral density
\begin{equation}\label{E3}
J(\omega)=\frac{1}{2\pi}\frac{\gamma_{0}\lambda^2}{(\omega_{0}-\omega)^2+\lambda^2},
\end{equation}
the parameter $\lambda$ defines the spectral width of the coupling which is associated with the reservoir correction time
by the relation $\tau_{E}=\lambda^{-1}$, and the parameter $\gamma$ is related to the relaxation
time scale by the relation $\tau_{R}=\gamma_{0}^{-1}$. In this sense, the boundary between Markovian regimes and non-Markovian
regimes can be approximately specified by the reservoir's parameters $\gamma_{0}$ and $\lambda$~\cite{Li2,Breuer}. Usually, $\gamma_{0} < \lambda /2$ $(\gamma_{0}/\lambda < \frac{1} {2 })$ represents the weak system-reservoir coupling regime where the behavior of the
system is the Markovian dynamic, and $\gamma_{0} > \lambda /2$ $(\gamma_{0}/\lambda > \frac{1} {2 })$ corresponds to the strong system-reservoir coupling regime where  the non-Markovian dynamic occurs.

With the above equations in mind, using Laplace transform and its inverse transform, the corresponding probability amplitude $C_{i}(t)$ is easily obtained
\begin{equation}\label{E4}
C_{i}(t)=\frac{N-1}{N}+\frac{\emph{e}^\frac{-\lambda t}{2}}{N}\left[\cosh(\frac{D t}{2})+\frac{\lambda}{D}\sinh(\frac{D t}{2})\right],
\end{equation}
with $D=\sqrt{\lambda_{}{}^2-2N\gamma_{0}\lambda}$. An interesting result for the above equation at the asymptotical
limit $t\rightarrow\infty$ reduces to $\frac{N-1}{N}$. On the other hand, as $N\rightarrow\infty$, $C_{i}(t\rightarrow\infty, N\rightarrow\infty)$ tends to 1.  In the following, we are going to investigate how the addition of qubits influences on QFI and GF of a single qubit system in a common dissipative reservoir.

\section{QFI preservation in a common dissipative reservoir through the additional qubits}
Before investigate the QFI dynamics, we firstly briefly review the main aspects of QFI which  is determined by the well-known Cram¨¦r-Rao inequality~\cite{2a,2b}
\begin{equation}\label{E5}
(\Delta\theta)^2\geq\frac{1}{n QFI_{\theta}}
\end{equation}
where $(\Delta\theta)^2$ represents the mean square error of the parameter $\theta$ , and $n$ denotes the number of repeated independent experiments.
$QFI_{\theta}$ is characterized the sensitivity of the state with respect to changes in a parameter, and it is defined as~\cite{2a,2b}
\begin{equation}\label{E6}
QFI_{\theta}=Tr[\rho(\theta)L^2],
\end{equation}
where $\rho(\theta)$ is the density matrix of the system,  and $L$ is the symmetric logarithmic derivation given by
\begin{equation}\label{E7}
\frac{\partial \rho(\theta)}{\partial \theta}=\frac{L\rho(\theta)+\rho(\theta)L}{2}.
\end{equation}
Making use of the spectrum decomposition $\rho(\theta)=\sum_{i}\lambda_{i}|\omega_{i}\rangle \langle \omega_{i}|$, analytical expression of QFI can be divided into two parts
\begin{equation}\label{E8}
QFI_{\theta}=\sum_{i}\frac{(\partial_{\theta}\lambda_{i})^2}{\lambda_{i}}+2\sum_{i,i'}\frac{(\lambda_{i}-\lambda_{i'})^2}{\lambda_{i}+\lambda_{i'}}|\langle \omega_{i}|\partial_{\theta}\omega_{i'}\rangle|^2,
\end{equation}
The purpose of the parameter estimate for a quantum system is to find the values of unknown parameter as precisely as possible, one should optimize input initial state to maximize the QFI. Based on the present model in this paper, the QFI based parameter is assumed to be induced by a phase gate $U(\theta)=|0\rangle \langle 0|+\exp(i\theta)|1\rangle \langle 1|$ acting on a system of interest. Then the interest system with additional qubits in a common reservoir evolve with time before the measurement performed on. In order to estimate the unknown parameter $\theta$ as precisely as possible, the optimal input state of the interest system may be chosen as $|\Psi_{opt}\rangle=\frac{1}{\sqrt{2}}(|0\rangle +|1\rangle)$, which maximizes the $QFI_{\theta}$ of the output state $U(\theta)|\Psi_{opt}\rangle$~\cite{Sun,Dorner}. Taking the output state $U(\theta)|\Psi_{opt}\rangle$ and the Eq. (2), the reduced density matrix $\rho(\theta)$ of the interest system is given by
\begin{equation}\rho_{}(\theta)=\left(
\begin{array}{ c c c c l r }
|C_{1}(t)|^2 & \frac{1}{\sqrt{2}}\exp(i\theta)C_{i}(t)\\
\frac{1}{\sqrt{2}}\exp(-i\theta)C_{i}^{\ast}(t) & 1-|C_{i}(t)|^2\\
\end{array}
\right).
\end{equation}

According to Eq.~(\ref{E8}), the analytic expression of $QFI_{\theta}$ of phase estimation for a single qubit system in a reservoir in the presence of additional qubits is obtained
\begin{equation}\label{E9}
QFI_{\theta}=\left\{\frac{N-1}{N}+\frac{\emph{e}^\frac{-\lambda t}{2}}{N}\left[\cosh(\frac{D t}{2})+\frac{\lambda}{D}\sinh(\frac{D t}{2})\right]\right\}^2,
\end{equation}
form above equation, an interesting result for $QFI_{\theta}$ occurs at the asymptotical limit $t\rightarrow\infty$, as follows
\begin{equation}\label{E10}
QFI_{\theta}=\left\{\frac{N-1}{N}\right\}^2,
\end{equation}
which is only dependent of the total number of qubits $N$ in a common reservoir. Especially, as $N\rightarrow\infty$, Eq.~(\ref{E10}) reduces to the maximal initial $QFI_{\theta}=1$.
\begin{figure}[htb]\centering
\includegraphics[width=6.5cm]{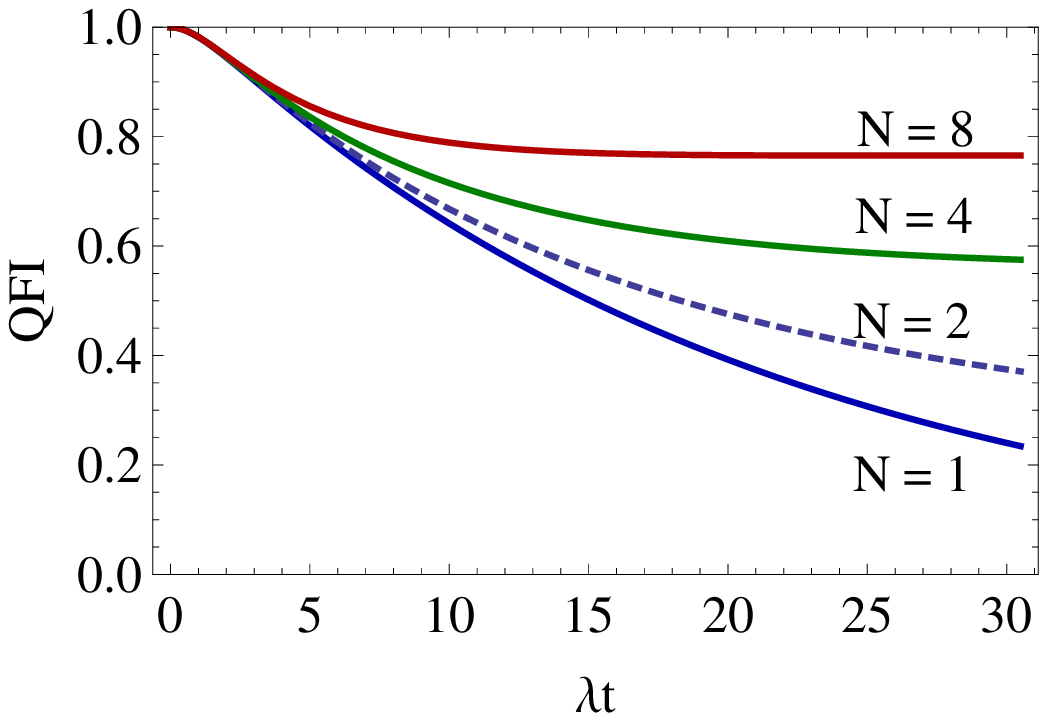}\quad
\caption{\label{Fig1}(Color online) QFI as a function of $\lambda t$ with the optimal input state $|\Psi_{opt}\rangle=\frac{1}{\sqrt{2}}(|0\rangle +|1\rangle)$ in Markovian regime $\gamma =0.05 \lambda $ for different number of qubits $N$}
\end{figure}
\begin{figure}[htb]\centering
\includegraphics[width=6.5cm]{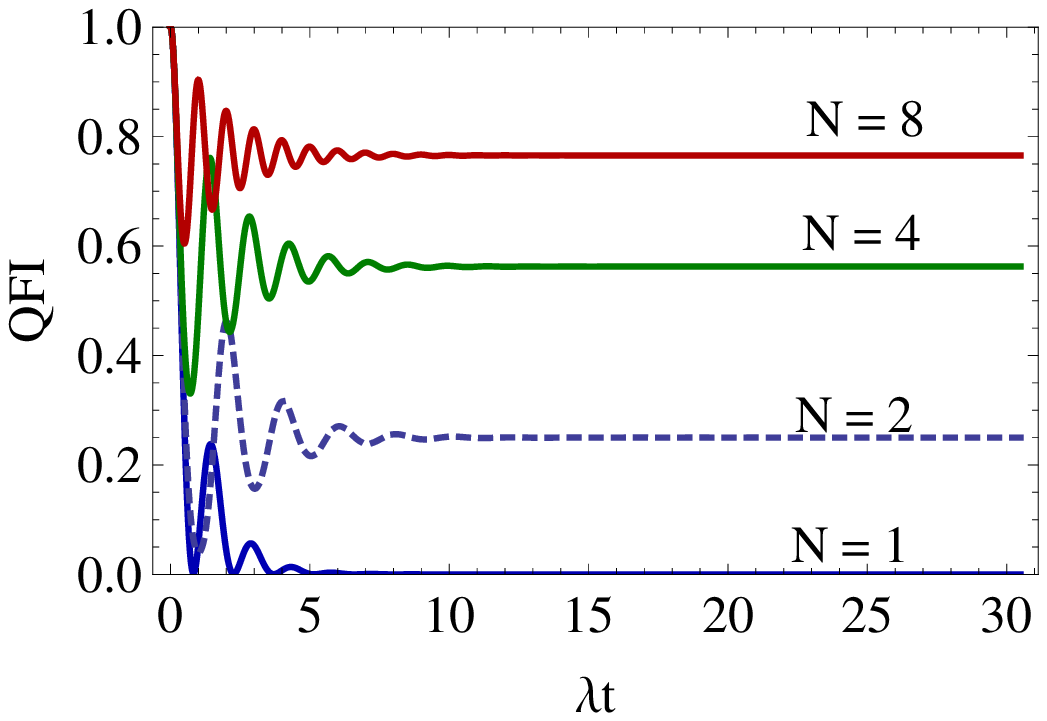}\quad
\caption{\label{Fig2}(Color online) QFI as a function of $\lambda t$ with the optimal input state $|\Psi_{opt}\rangle=\frac{1}{\sqrt{2}}(|0\rangle +|1\rangle)$ in non-Markovian regime $\gamma =10\lambda$ for different number of qubits $N$.}
\end{figure}

\begin{figure}[htb]\centering
\includegraphics[width=6.5cm]{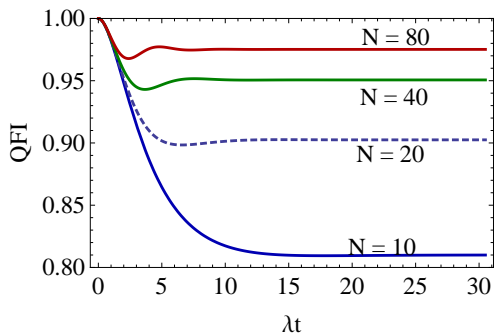}\quad
\caption{\label{Fig3}(Color online) The effects of the number of additional qubits $N-1$ on the QFI in non-Markovian regime $\gamma =0.05\lambda$.}
\end{figure}
In the following, we will explore the influence of the number of additional qubits  on the QFI, and compare the precision of  parameter estimation in both weak and strong coupling regimes.
Fig. 1 shows QFI dynamics of a single-qubit system as a function of $\lambda t$ for different the number of additional qubits in the weak coupling regime $\gamma_{0} =0.05 \lambda$.  As $N=1$(in the absent of additional qubits case), namely, a two-level (qubit) system is coupling to a zero-temperature bosonic reservoir, the dynamics of QFI had been studied in Refs.~\cite{Berrada1,Sun}, and found the dynamics of QFI monotonically decays to zero without revivals.  However, with the number of additional qubits  increasing, (e.g. $N=2, N=4$, and $N=8$), the dynamics of QFI shows a low speed of decay as the system evolves. Notice that the QFI does not decay to zero during the time evolution in the present of addition qubits, but it goes down to a stead value which is dependent on the number of additional qubits $N-1$ (or the total number of qubits $N$). One can easily determine from Eq.~(\ref{E9}) at the limited $t\rightarrow \infty$ which reduces to $\frac{(N-1)^2}{N^2}$.  In particularly, $N\rightarrow\infty$, the maximal QFI approaches to 1. In contrast to QFI in the weak coupling regime, Fig. 2 shows QFI dynamics in the strong coupling regime $\gamma_{0} =10 \lambda$. It is observed that, for without addition qubits $N=1$, the dynamics of QFI non-monotonically decays, but experiences damped revival oscillations. This result is consistent with the previous analysis in Refs.~\cite{Berrada1,Sun}. As the number of additional qubits  goes up, QFI dynamics displays more quickly damped revival oscillations and stays at some steady value. The larger the number of additional qubits is, the smaller the amplitude for QFI oscillates. In particularly, $N\rightarrow\infty$, the maximal QFI approaches to 1. By comparing QFI in the weak coupling regime (Fig.1) with the strong coupling regime (Fig. 2), we find that, adding qubits into environmental reservoir can dramatically protect the QFI so as to enhance the precision of parameter estimation. This result indicates QFI of dissipation can be suppressed by
increasing number of additional qubits. The more the number of additional qubits are, the much more slowly the QFI decays, the higher the estimation precision.

It is worth pointing out that the system's dynamics of a two-level system coupling to $N$ bosonic reservoirs has been reported by Man et al~\cite{Man}, and they found that the system's dynamics was strongly depended on reservoirs' number $N$. The system dynamics remains Markovian when $1\leq N < N_{c}$ ($N_{c}=[\lambda_{n}/2\gamma_{n} +1]\in Z$), but $N \geq N_{c}$ corresponds to non-Markvoian dynamics. However, different from the model by Man proposed, we here consider $N$ qubits are immersed in a common reservoir, each of qubit has $\gamma_{n} < \lambda_{n} /2$,  the system's dynamics varies with $N$(the total qubits number). To get a better understanding of the effects of the number of additional qubits  on the QFI,  Fig.3 shows QFI dynamics in the weak coupling regime $\gamma_{0} =0.05 \lambda$. One can find that, on one hand, QFI displays monotonically decays without revivals for
$1\leq N < N_{c}$($ N_{c}=11$). On the other hand, QFI dynamics displays damped revival oscillations for $N \geq 11$ even in this weak coupling regime. This physical origin  can understand as followings. When $N$ identical qubits are all independently coupling to a common dissipative reservoir in the weak coupling regime $\gamma_{n} < \lambda_{n} /2$ $(\frac{\gamma_{n}}{\lambda_{n}} < \frac{1} {2 })$, then $\sum_{n=1}^{N}\frac{\gamma_{n}}{\lambda_{n}}>\frac{1} {2 }$ which is identical to the strong system-reservoir coupling regime. Therefore, the QFI dynamics is non-Markovian. This result indicates $N_{c}$ can be regarded
as a critical number of qubits no matter how strong the couplings between the individual qubit and the reservoir.

\section{The GP of a single-qubit system in the presence of additional qubits}
In this section,  we study quantitatively the GP of a single qubit system along with $N-1$ qubits coupled to a common dissipative reservoir in terms of additional qubits.  We apply the approach proposed in Ref.~\cite{Tong}, where the authors developed a kinematic approach to GP for open quantum systems in non-unitary evolution induced by dissipative environments. GP is defined as
\begin{eqnarray}\label{Eq13}
\Phi_{GP}&=&Arg\{\sum_{i}\sqrt{\lambda_{i}(0)\lambda_{i}(t)}\langle\omega_{i}(0)|\omega_{i}(t)\rangle\nonumber \\
&\times&\exp{[-\int_{0}^{t}\langle\omega_{i}(\tau)|\frac{d}{d\tau}\omega_{i}(\tau)\rangle d\tau}]\}
\end{eqnarray}
where $\lambda_{i}(t)$ and $\omega_{i}(t)$ are the eigenvalues and eigenvectors of the matrix $\rho(t)$ given by Eq.~(\ref{E1}), and Arg denotes the argument of the complex number.  To demonstrate the influence of the addition qubits  and the initial conditions of the system on the GP, we assume that the initial state of the interest system is chosen as
$|\Psi\rangle=\cos\frac{\theta}{2}|0\rangle +\sin\frac{\theta}{2}|1\rangle$. Substituting Eq.~(\ref{E1}) and Eq.~(\ref{E4}) into Eq.~(\ref{Eq13}), then the GP of the
qubit acquired after a quasicyclic $t = 2\pi/\omega_{0}$ can be calculated as
\begin{eqnarray}
\label{eq1}
\Phi_{GP}=\int_{0}^{t}\frac{4(|C_{i}(t)|^2\cos^2\frac{\theta}{2}-\lambda)^2}{|C_{i}(t)|^2\sin^2\theta+4(|C_{i}(t)|^2\cos^2\frac{\theta}{2}-\lambda)^2}dt,
\end{eqnarray}
in which $\lambda=\frac{1}{2}\left[1-\sqrt{|C_{i}(t)|^2\sin^2\theta+(2|C_{i}(t)|^2\cos^2\frac{\theta}{2}-1)^2}\right]$.
\begin{figure}[htb]\centering
\includegraphics[width=6.5cm]{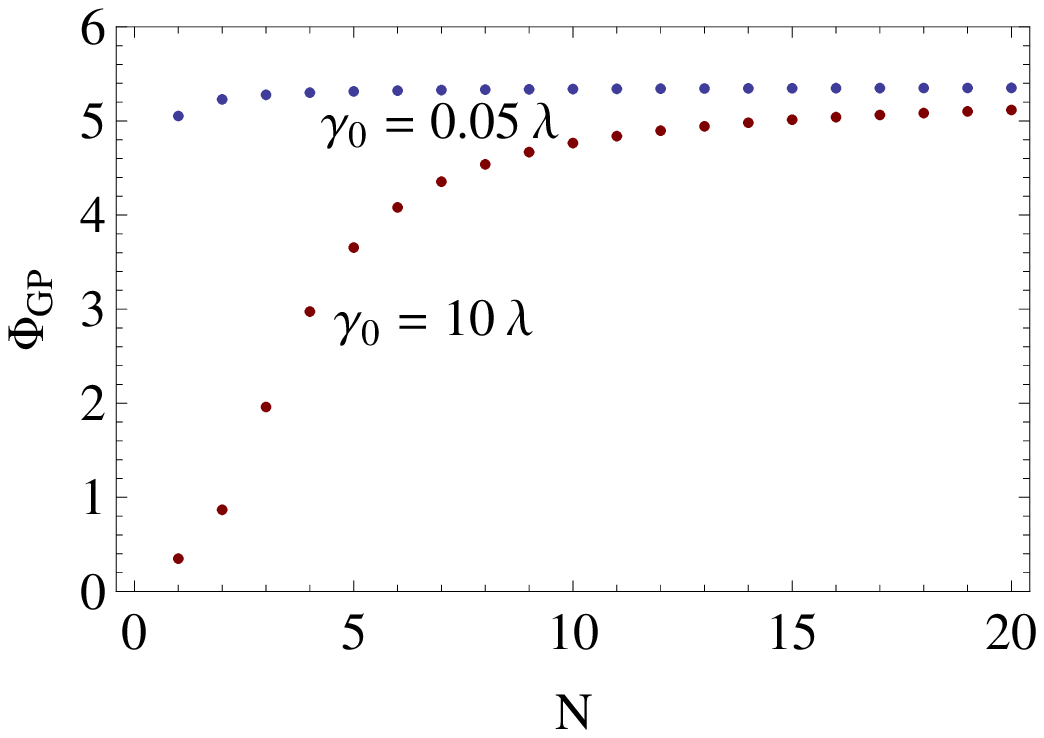}\quad
\caption{\label{Fig4}(Color online) GP as a function of $N$ in Markovian regime $\gamma_{0} =0.05 \lambda $ and non-Markovian regime $\gamma_{0} =10 \lambda $.}
\end{figure}

\begin{figure}[htb]\centering
\includegraphics[width=6.5cm]{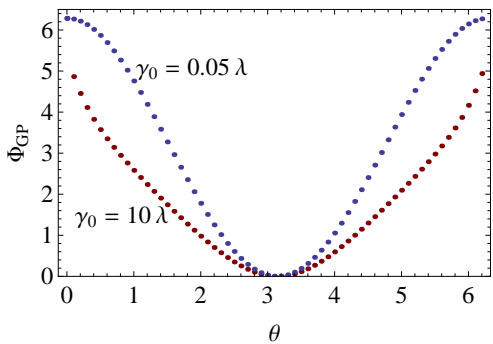}\quad
\caption{\label{Fig5}(Color online) GP as a function of $\theta$ in Markovian regime $\gamma_{0} =0.05 \lambda $ and non-Markovian regime $\gamma_{0} =10 \lambda $.}
\end{figure}
First, we concentrate on the influence of the addition number of qubits on GP, we display the GP as a function of $N$ for fixed the initial state's parameter $\theta=\pi/4$ in the Markovian regime $\gamma_{0} =0.05 \lambda$ and non-Markovian regime $\gamma-{0} =10 \lambda$ in Fig.4, respectively. It is obvious to see the values of GPs firstly raise monotonically with the increase of the addition number of qubits and finally reach their maximal values in both Markovian regime and non-Markovian regime. The saturation maximal values is only related to the initial state's parameter $\theta$. One can easily determine from Eq.~(\ref{eq1}) which reduces to $\Phi_{GP}^{0}=\pi(1+\cos\theta)$ when the addition number of qubits is large enough. It is worthy noting that, the saturation maximal value $\Phi_{GP}^{0}$ is just the GP acquired by a single-qubit
in the unitary dynamics. This is consistent with the result obtained in Ref.~\cite{Tong,Yi1}. However, Such monotonic increasing behaviors of GPs in both Markovian regime and non-Markovian regime are dramatically affected by the additional qubits, in particular, the number of additional qubits is small. However, when the number of additional qubits is large enough, the GPs are immune to the decoherence of environment in both Markovian and non-Markovian regimes. From the above results, we may make a control on the robustness of GP of the system against the dissipative environment by increasing the number of qubits in a dissipative reservoir. Apart from the effects of the addition of qubits, the initial state's parameter of the system also play an important role in the evolution of GP. In Fig. 5, we plot the GP as a function
of the initial state's parameter $\theta$ for $N=2$ in Markovian regime $\gamma_{0} =0.05 \lambda $ and non-Markovian regime $\gamma_{0} =10 \lambda$, respectively. One can see that, the GP first deceases and then increases, vanishes for $\theta=\pi$.  They are symmetric in a period $[0,2\pi]$. This results reflect the GPs are sensitive to the initial state's parameter.

\section{Conclusion}

Though some recent works are based on the environment effects on the QFI dynamics of the open system in the weak coupling regime~\cite{Berrada1,Sun}, there are still some interesting results that need further discussion and in-depth study. In this paper, we study the QFI for a single-qubit system coupled to a common dissipative reservoirs through the addition qubits into reservoir. We mainly focus on the influence of the additional noninteracting qubits on the QFI dynamics. The results show that, the QFI is completely protected in both Markovian and non-Markovian regimes by increasing the number of additional qubits. Besides, the phenomena of QFI display monotonic decay or non-monotonic with revival oscillations depending on the number of additional qubits $N$ in a common dissipative reservoir. If $N < N_{c}$, the behavior of QFI with monotonic decay occurs. However, if $N \geq N_{c}$, QFI exhibits non-monotonic behavior with revival oscillations for whatever system-reservoir coupling strengths. Moreover, we extend this model to investigate the effect of the addition of qubits and the initial conditions of the system on the GP. The GP for a nonunitary evolving qubit is calculated according to the formula of Ref.~\cite{Tong}, we find that, the GP is unaffected by the dissipative reservoir with the number of additional qubits increasing, regardless of whether the qubit system is in Markovian or non-Markovian regimes. However, the GP is sensitive to the initial parameter $\theta$, and possesses symmetric in a range regime $[0,2\pi]$.

\acknowledgments
This research is supported by the Start-up Funds for Talent Introduction and Scientific Research of Changsha University 2015 (Grant No.SF1504), the Scientific Research Project of Hunan Province Department of Education (Grant Nos.16C0134 and 17C0133), the Natural Science Foundation of Hunan Province (Grant No.2017JJ3346), Key Laboratory of Low-Dimensional Quantum Structures and
Quantum Control of Ministry of Education (QSQC1403) and the Project of Science and Technology Plan of Changsha (ZD1601071)

\label{app:eff-trans}

\end{document}